\documentclass[conference,a4paper]{IEEEtran}

\pdfoutput=1

\usepackage{xcolor}
\usepackage{balance}
\usepackage{comment}
\usepackage{amsmath}
\usepackage{amssymb}
\usepackage{layouts,flushend}

\ifCLASSINFOpdf
  \usepackage[pdftex]{graphicx}
\else
  \usepackage[dvips]{graphicx}
\fi

\ifCLASSOPTIONcompsoc
 \usepackage[caption=false,font=normalsize,labelfont=sf,textfont=sf]{subfig}
\else
 \usepackage[caption=false,font=footnotesize]{subfig}
\fi

\begin{document}
\title{$1$-Bit SubTHz RIS with Planar Tightly Coupled Dipoles: Beam Shaping and Prototypes}
\author{\IEEEauthorblockN{
Xianjun Ma\IEEEauthorrefmark{1}, Yonggang Zhou\IEEEauthorrefmark{1}, Qi Luo\IEEEauthorrefmark{2}, Yihan Ma\IEEEauthorrefmark{2}, Kyriakos Stylianopoulos\IEEEauthorrefmark{3}, and George C. Alexandropoulos\IEEEauthorrefmark{3}  
}                                    
\IEEEauthorblockA{\IEEEauthorrefmark{1}
Nanjing University of Aeronautics and Astronautics, Nanjing, China}
\IEEEauthorblockA{\IEEEauthorrefmark{2}
University of Hertfordshire, Hatfield, UK}
\IEEEauthorblockA{\IEEEauthorrefmark{3}
Department of Informatics and Telecommunications, National and Kapodistrian University of Athens, Greece}
\IEEEauthorblockA{emails: zyg405@nuaa.edu.cn, \{q.luo2, y.ma20\}@herts.ac.uk, \{kstylianop, alexandg\}@di.uoa.gr}
}

\maketitle
\begin{abstract}
In this paper, a proof-of-concept study of a $1$-bit wideband reconfigurable intelligent surface (RIS) comprising planar tightly coupled dipoles (PTCD) is presented. The developed RIS operates at subTHz frequencies and a $3$-dB gain bandwidth of $27.4\%$ with the center frequency at $102$ GHz is shown to be obtainable via full-wave electromagnetic simulations. The binary phase shift offered by each RIS unit element is enabled by changing the polarization of the reflected wave by $180^\circ$. The proposed PTCD-based RIS has a planar configuration with one dielectric layer bonded to a ground plane, and hence, it can be fabricated by using cost-effective printed circuit board (PCB) technology. We analytically calculate the response of the entire designed RIS and showcase that a good agreement between that result and equivalent full-wave simulations is obtained. To efficiently compute the $1$-bit RIS response for different pointing directions, thus, designing a directive beam codebook, we devise a fast approximate beamforming optimization approach, which is compared with time-consuming full-wave simulations. Finally, to prove our concept, we present several passive prototypes with frozen beams for the proposed $1$-bit wideband RIS. 
\end{abstract}

\vskip0.5\baselineskip

\section{Introduction}
Reconfigurable intelligent surfaces (RISs) have been identified as one of the key physical layer technologies for the upcoming sixth generation (6G) of wireless communications~\cite{ETSI,WavePropTCCN,Tsinghua_RIS_Tutorial}. Such metasurfaces are capable of reflecting their incident waves to desired angles, enriching specular channels with scattering, and enabling over-the-air analogue computations~\cite{RIS_space_shift_keying,alexandg_2021}, paving the way for the envisioned smart wireless environments~\cite{Strinati_2021a,EURASIP_RIS_all}. Another highly expected feature of 6G is the promising utilization of the THz frequency band that encapsulates extensive unallocated bandwidth, which can be used to support rate- and localization/sensing-demanding applications, while exhibiting increased confidentiality and anti-interference capabilities~\cite{9794668}. 

The combination of the latter two 6G technologies is lately attracting significant research~\cite{TERRAMETA_website,Keykhosravi2022infeasible}, mainly due to the fact that reconfigurable metasurfaces can contribute in extending the coverage of THz links. RIS unit cells operating at THz were recently presented in~\cite{THz_RIS_1,THz_RIS_2,THz_RIS_3}, where the tunable characteristics of their reflective beams were achieved through a tunable dielectric substrate equipped with voltage control. In~\cite{THz_RIS_4}, the beam tuning characteristics were dynamically controlled based on a liquid crystal unit structure, indicating that this design method of reconfigurable reflection arrays is a promising one for THz RISs. A broadband tightly coupled reflectarray structure has been also reported for THz RIS unit elements in~\cite{THz_RIS_5,wang_efficiency_2020}. According to the relevant literature, the tightly coupled dipole technology has the advantage of being ultra wideband for antenna application, which makes tightly coupled antenna elements another candidate design solution for THz RISs. In fact, remarkable progress has been made in the design of tightly coupled RIS unit elements for those frequencies, and $1$-bit and $2$-bit wideband and ultra-wideband reflectarrays have been designed~\cite{kamoda_60-ghz_2011,han_wideband_2019,luo_ultra-wideband_2019}.

In this paper, we capitalize on our recent ultra-wideband subTHz RIS~\cite{Luo_THz_RIS}, which was based on tightly coupled dipoles that are vertically placed on a ground plane, and present a novel efficient RIS unit element design. Although the elements in the design of \cite{Luo_THz_RIS} can be fabricated by using conventional printed circuit board (PCB) technology, the assembly of each individual RIS element is challenging and the mechanical robustness of the overall design is poor. We have thus improved that design by modifying the RIS unit cell to a planar structure. This design can be easily fabricated via multi-layer PCB technology with significantly reduced complexity and cost for the assembly process. Compared with the design in~\cite{Luo_THz_RIS}, the bandwidth of the presented subTHZ RIS design is reduced, but the resulting planar tightly coupled dipoles (PTCD)-based RIS still possesses a wideband response. In particular, the simulated $3$-dB gain bandwidth is $27.4\%$ with the center frequency at $102$ GHz. 

The remainder of the paper is organized as follows. Section~II details the design of the proposed RIS unit cell, including full-wave simulations for the entire RIS. Section~III presents the beamforming modeling and algorithmic approach to calculate the response of the developed $1$-bit wideband RIS to arbitrary directions, while Section~IV concludes the paper.

\section{The Proposed PTCD-based subTHz RIS Design}\label{sec:ris}
In this section, we present the configuration of the planar tightly coupled dipole, which is used as the unit element of the proposed subTHz RIS. Simulated radiation patterns via electromagnetic (EM) software of the entire RIS are also included and discussed.

\subsection{$1$-Bit Unit Cell Design}\label{element_design}
\begin{figure}[!t]
\centering
\includegraphics[width=\columnwidth]{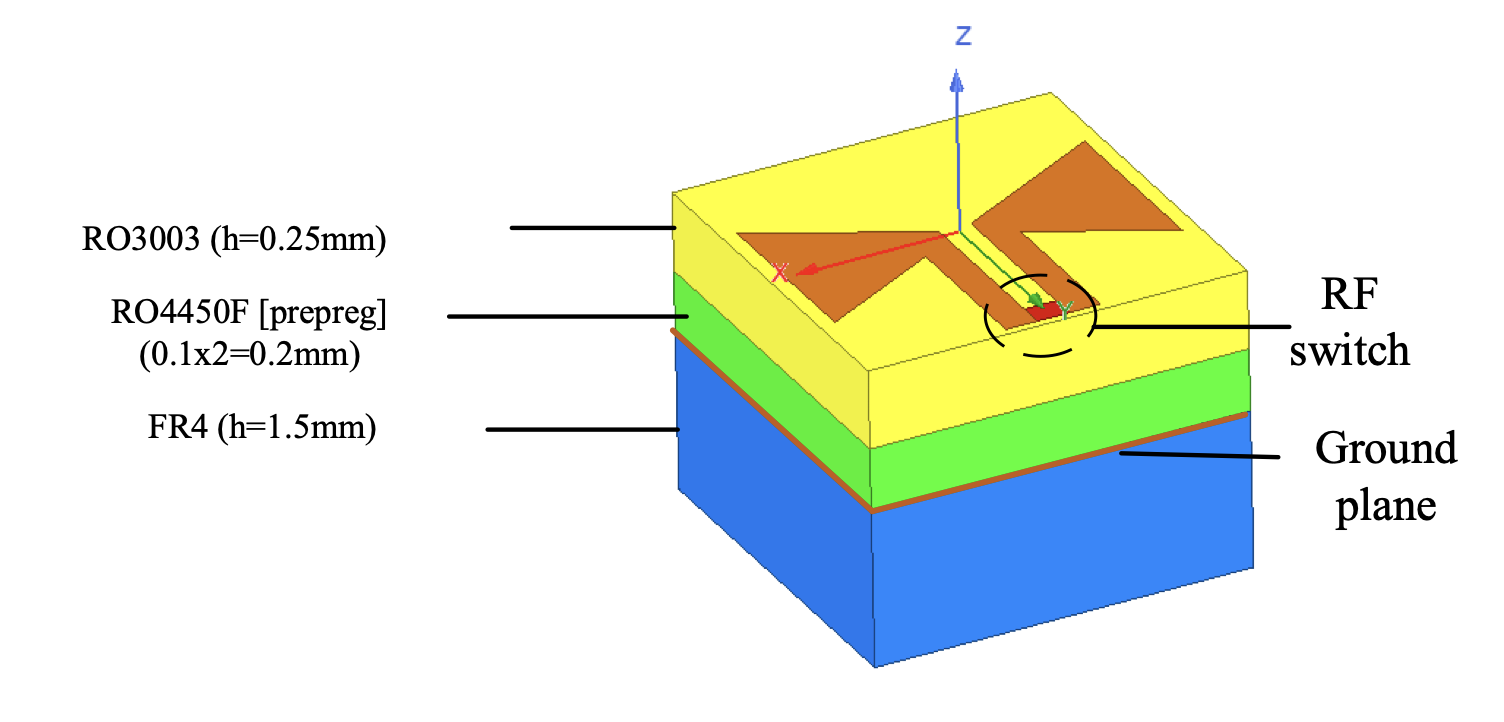}
\caption{The illustration in 3D of the designed RIS unit cell, which is a planar tightly coupled butterfly dipole printed on the top layers of the substrate. The RIS element is linearly polarized in the $x$-axis. An RF switch is placed in the middle of the dipole to rotate the polarization of the impinging wave.}
\label{fig:figure_1}
\end{figure}

Fig.~\ref{fig:figure_1} illustrates the configuration of the proposed PTCD structure which constitutes the unit cell of the designed subTHz RIS. The unit cell is designed with a printed butterfly dipole element placed on a RO3003 substrate with a thickness of $0.25$ mm. Below the RO3003, there is a RO4450F substrate with a thickness of $0.2$ mm, which is also backed by a ground plane on the other side. Additionally, another dielectric layer, FR4 with a thickness of $1.5$ mm, is introduced below the metallic ground to enhance the overall structure's mechanical strength. As a result, the total thickness of the reflecting unit cell is $2$ mm, including the thickness of the copper layers. By controlling the ON/OFF states of the RF switch placed at the same layer of the printed dipole, the reflection phase of the unit cell can be tuned with $180^\circ$ phase difference. As a proof-of-concept study, at this stage, we model the RF switch as an ideal one and do not consider the equivalent circuit model of the switch, which is left for the extension of this work. 

\begin{figure}[!t]
\centering
\includegraphics[width=0.5\columnwidth]{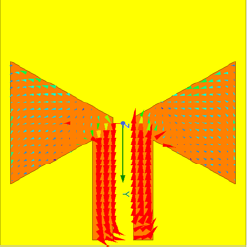}
\caption{The current distribution of the designed $1$-bit RIS unit cell when the RF switch is ON.}
\label{fig:figure_2}
\end{figure}

\begin{figure}[!t]
\centering
\includegraphics[width=0.5\columnwidth]{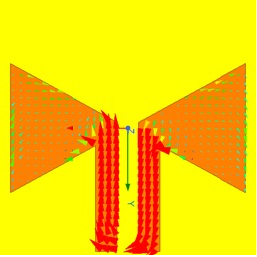}
\caption{The current distribution of the designed $1$-bit RIS unit cell when the RF switch is OFF.}
\label{fig:figure_3}
\end{figure}

\begin{figure}[!t]
\centering
\includegraphics[width=\columnwidth]{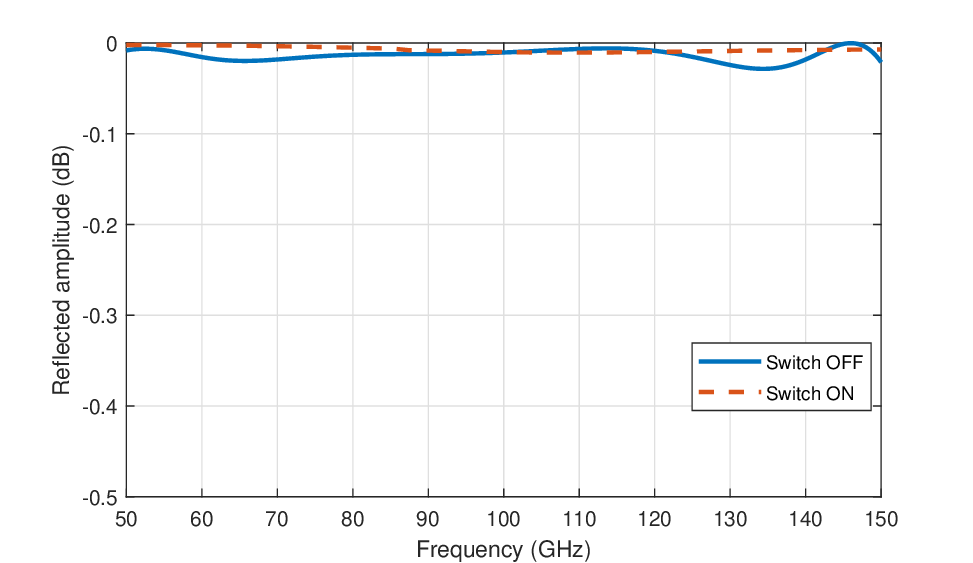}
\caption{The reflection amplitude of the designed $1$-bit RIS unit cell when the RF switch is ON and OFF as a function of the operating frequency in the $[88.5, 116.7]$ GHz range.}
\label{fig:figure_4}
\end{figure}

\begin{figure}[!t]
\centering
\includegraphics[width=\columnwidth]{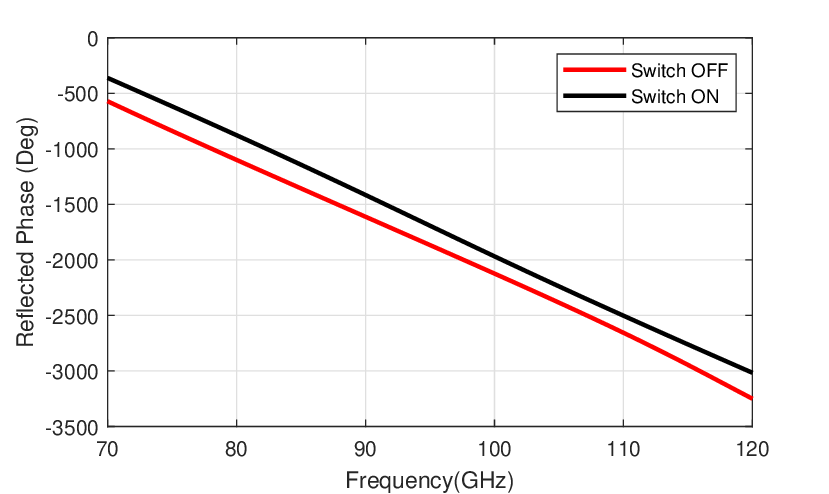}
\caption{The reflection phase of the designed $1$-bit RIS unit cell when the RF switch is ON and OFF as a function of the operating frequency in the $[88.5, 116.7]$ GHz range.}
\label{fig:figure_5}
\end{figure}

The current distribution of the dipole for different states of the RF switch is depicted in Figs.~\ref{fig:figure_2} and~\ref{fig:figure_3}. As shown, the surface current on the planar dipole was rotated by $180^o$, implying that the polarization of the reflected wave is changed by $180^o$. In this simulation, a linear polarized normal incident wave along the $x$ direction is employed to illuminate the dipole antenna with the same linear polarization. Figs.~\ref{fig:figure_4} and~\ref{fig:figure_5} demonstrate the corresponding reflection phase and amplitude of the unit cell with binary phase states, considering the normal incident angle. As shown in Fig.~\ref{fig:figure_4}, the RIS unit cell shows low insertion loss. It is noted that the actual insertion loss of the RIS unit cell will be higher when considering the equivalent circuit model of the subTHz switch. From Fig.~\ref{fig:figure_5}, it can be seen that the unit cell exhibits a wideband phase response with stable $180^o$ phase differences from the frequency $88$ to $116$ GHz for the two different states of the RF switch. 

\subsection{RIS Design and Simulated Response}\label{RIS_performance}
\begin{figure}[!t]
\centering
\includegraphics[width=0.6\columnwidth]{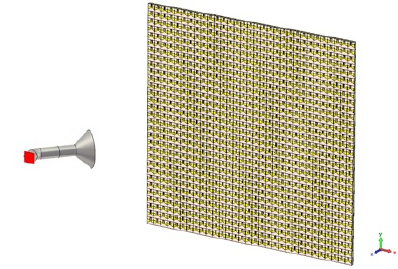}
\caption{The simulation model of the wideband subTHz RIS consisting of $31\times31$ tightly-coupled dipole elements.}
\label{fig:figure_6}
\end{figure}

We use the previously presented unit cell to synthesize an RIS aperture comprising $31\times31$ elements in CST Microwave Studio. Considering the limitation of the testing facility, we choose to place the source of the incident wave relatively close to the RIS, and in particular, we set $f/D=0.5$, where $f$ is the focal distance and $D$ is the diameter of the aperture of the reflecting surface. The simulation model for the designed RIS is illustrated in Fig.~\ref{fig:figure_6}. 

\begin{figure}[!t]
\centering
\includegraphics[width=\columnwidth]{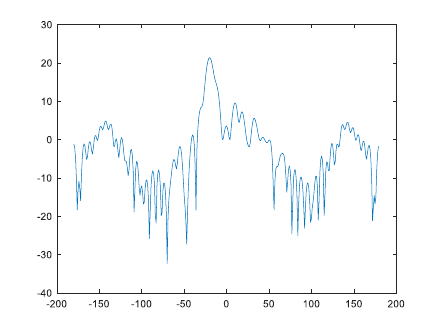}
\caption{The simulated radiation of the designed wideband subTHz RIS when the beam is configured to point to the $22.5^\circ$ direction.}
\label{fig:figure_9}
\end{figure}

In Fig.~\ref{fig:figure_9}, the simulated radiation pattern of the designed RIS at the $H$-plane is demonstrated when the reflection beam is configured to $22.5^\circ$. It is noted that despite the phase error resulting from the $1$-bit phase quantization, the beam pointing error is small and the sidelobe level (SLL) is larger than $10$ dB. We have also conducted full-wave EM simulations to extract the active radiation pattern of the unit cell and analytically calculated the response of the entire RIS. Figs.~\ref{fig:figure_10} and~\ref{fig:figure_11} depict the results of the pattern calculated by the active method and that of equivalent EM simulations, when the RIS is reconfigured to $0^\circ$ and $22.5^\circ$ beam scanning, respectively. The approximated beams will be discussed next in Section~\ref{sec:beam_opt}. As observed, the analytical results agree well with the simulated results for the main beam, while some differences are present in the SLLs. The disagreement is mainly caused due to the fact that, during the analytical calculation, we ignored the change of the mutual couplings at the edge elements and the response of the RIS unit cell with arbitrary incident angles. According to the simulation results, it was found that when the RIS reflects the incident wave to foresight, the $3$-dB gain bandwidth is $27.4\%$ with the center frequency at $102$ GHz.

\begin{figure}[!t]
\centering
\includegraphics[width=\columnwidth]{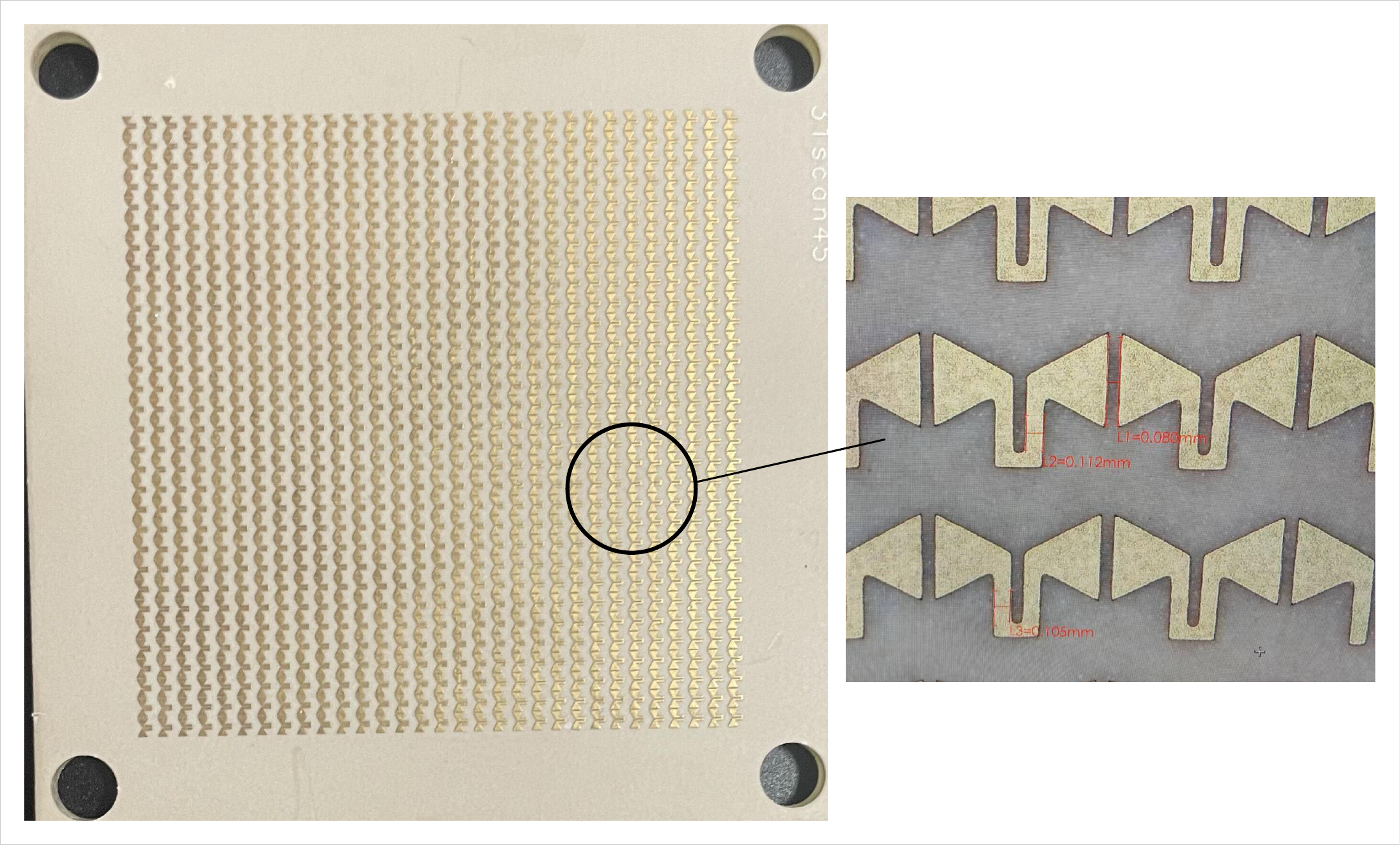}
\caption{Photo of the fabricated $31\times31$ subTHZ RIS with $1$-bit unit cells.}
\label{fig:prototype}
\end{figure}

\begin{figure}[!t]
\centering
\includegraphics[width=\columnwidth]{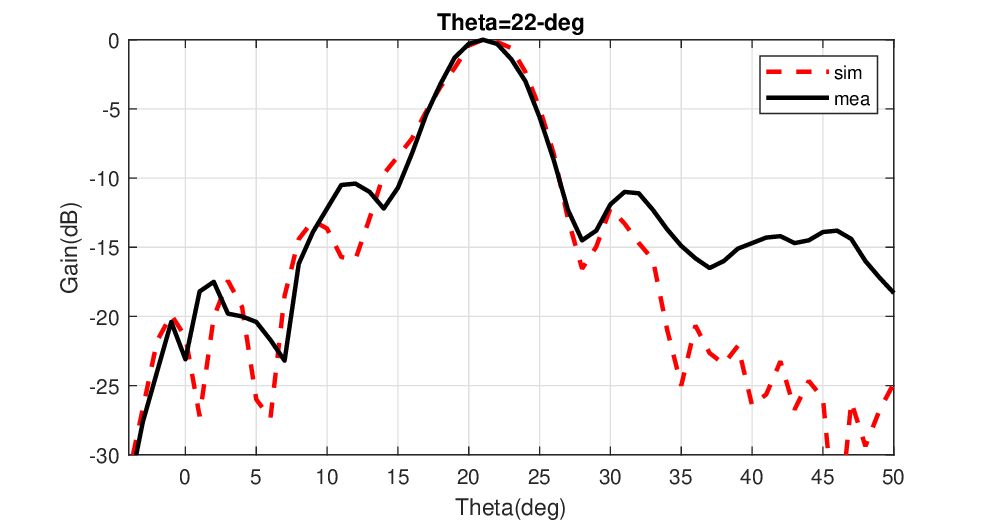}
\caption{The measured and simulated radiation of the designed wideband subTHz RIS when the beam is configured to point to the $22.5^\circ$ direction.}
\label{fig:RP_mea_22}
\end{figure}

To validate the proposed wideband subTHz RIS design with $1$-bit PTCD-based unit cells and evaluate the fabrication accuracy, several passive prototypes have been fabricated. These prototypes were designed to provide reflection beams pointing at the fixed angles $0^\circ$, $22.5^\circ$, and $45^\circ$. Fig.~\ref{fig:prototype} shows a photo of one of the fabricated prototypes. Fig.~\ref{fig:RP_mea_22} shows the comparison between the measured and simulated radiation pattern with the beam steered to $22.5^\circ$ degree. As shown, good agreement at the main lobe is obtained. 

\section{RIS Beam Shaping Optimization}\label{sec:beam_opt}
In this section, we capitalize on the RIS beamforming optimization approach recently developed in \cite{EURASIP_RIS_beams_all} to devise reflection beams for the proposed wideband PTCD-based $1$-bit RIS design to any desired direction. This approach can be quite efficient for efficiently designing static as well as dynamic RIS beam codebooks for communications~\cite{Jamali2022,RIS_Hierarchical,An_Xu_Codebook}, localization/sensing~\cite{Keykhosravi2022infeasible,locrxris_all}, as well integrated sensing and communications~\cite{RIS_ISAC_SPM}. 

To deploy the approach in \cite{EURASIP_RIS_beams_all}, we consider a system model comprising a single-antenna transmitter (TX) placed in front of the center of the broadside of the designed RIS as in Section~\ref{RIS_performance}, and a candidate receiver (RX) antenna at an intended position $\mathbf{p}$.
Considering the pure line of sight (LoS) reflected link from the TX to the RX, and given an arbitrary RIS configuration vector $\boldsymbol{\omega}^T$
$\in \mathbb{C}^{M\times 1}$ for an RIS of $M$ unit cells, the corresponding beam pattern toward $\mathbf{p}$ has the form:
\begin{equation}\label{eq:beam_pattern_model}
    G(\mathbf{p})=  \mathbf{\boldsymbol{\omega}}^{\rm{T}}\mathbf{b}(\mathbf{p},\mathbf{p}_{\rm TX}),  
\end{equation}
where $\boldsymbol{\omega}$ is an $M$-element column vector including the RIS beam pattern and $\rm{T}$ represents vector transposition, $\mathbf{p}_{\rm TX}$ is the TX position, $\mathbf{b}(\mathbf{p},\mathbf{p}_{\rm{TX}})=\mathbf{a}(\mathbf{p})\odot\mathbf{a}(\mathbf{p}_{\rm{TX}})$, with $\odot$ denoting the element-wise product, and $[\mathbf{a}(\mathbf{p})]_m = \exp\left(-j\frac{2\pi}{\lambda}\left( \|\mathbf{p}-\mathbf{p}_m \|-\|\mathbf{p}-\mathbf{p}_{\rm{RIS}}\|\right) \right)$ denotes the beam-focusing vector with $\lambda$ being the wavelength, $\mathbf{p}_m$ is the position of the $m$-th ($m=1,2,\dots M$) RIS unit cell, and $\mathbf{p}_{\rm{RIS}}$ represents the RIS center location. This formulation holds both for near- and far-field scenarios, however, for our specific setup, we choose the RX coordinates at a fixed distance well beyond the Fraunhofer limit to investigate beam patterns that depend only on the targeted RX's angles.
The positions of the TX, $\mathbf{p}_{\rm{TX}}$, and that of the RIS, $\mathbf{p}_{\rm{RIS}}$, are considered fixed.
According to \cite[Sec.~3.2]{EURASIP_RIS_beams_all},
the RIS beam pattern $\boldsymbol{\omega}$ in \eqref{eq:beam_pattern_model} can be optimized to realize any desired reflection beam for any given RX position, disregarding the environmental factors of path loss, transmission power, and RX noise. To this end, the beam pattern for a directional beam can be represented by:
\begin{equation}
    G(\mathbf{p}) = \left(\mathbf{b}^{\ast}\left(\mathbf{p},\mathbf{p}_{\rm TX}\right)\right)^{\rm{T}}\mathbf{b}\left(\mathbf{p},\mathbf{p}_{\rm TX}\right),
\end{equation}
where $\mathbf{b}^{\ast}\left(\mathbf{p},\mathbf{p}_{\rm TX}\right) = \boldsymbol{\omega}^T$ is the conjugate of $\mathbf{b}\left(\mathbf{p},\mathbf{p}_{\rm TX}\right)$.

Following this system modeling approach, the obtained phase shifts of the RIS configuration fall under the continuous regime in $[0, 2\pi]$. Then, using the derivations presented in \cite{EURASIP_RIS_beams_all}, a discretized beam pattern $\hat{G}(\mathbf{p})$ (i.e., a discrete approximation of $G(\mathbf{p})$) can be numerically computed, whose states are guaranteed to belong in an arbitrary discrete set $\mathcal{K}$. For the $1$-bit subTHz RIS design presented in the previous Section~\ref{RIS_performance}, the phase shift of each RIX unit cell can take values from the discrete set $\mathcal{K}=\{0,\pi \}$, while their reflection amplitudes can be safely assumed to be unitary.
The RIS beam-shaping methodology involves iteratively taking appropriately scaled gradient steps over a modified version of the unconstrained objective, before projecting onto the feasible set of solutions. The complete formulation and algorithmic approach for RIS beam shaping is given in Algorithm~$3$ of~\cite{EURASIP_RIS_beams_all}.
\begin{figure}[!t]
\centering
\includegraphics[width=\columnwidth]{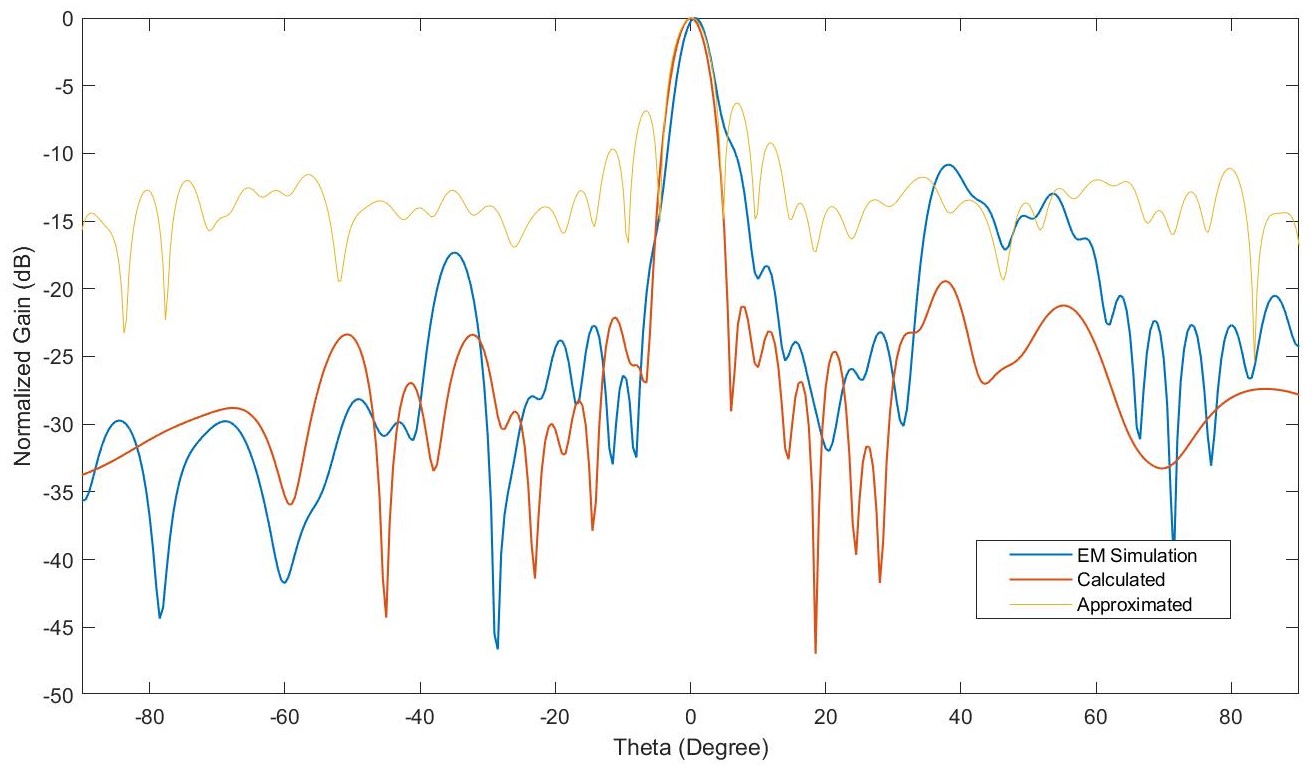}
\caption{Full-wave EM simulated, analytical, and approximated beam patterns when the RIS is configured to reflect the incident wave to the $0^\circ$ direction.}
\label{fig:figure_10}
\end{figure}
\begin{figure}[!t]
\centering
\includegraphics[width=\columnwidth]{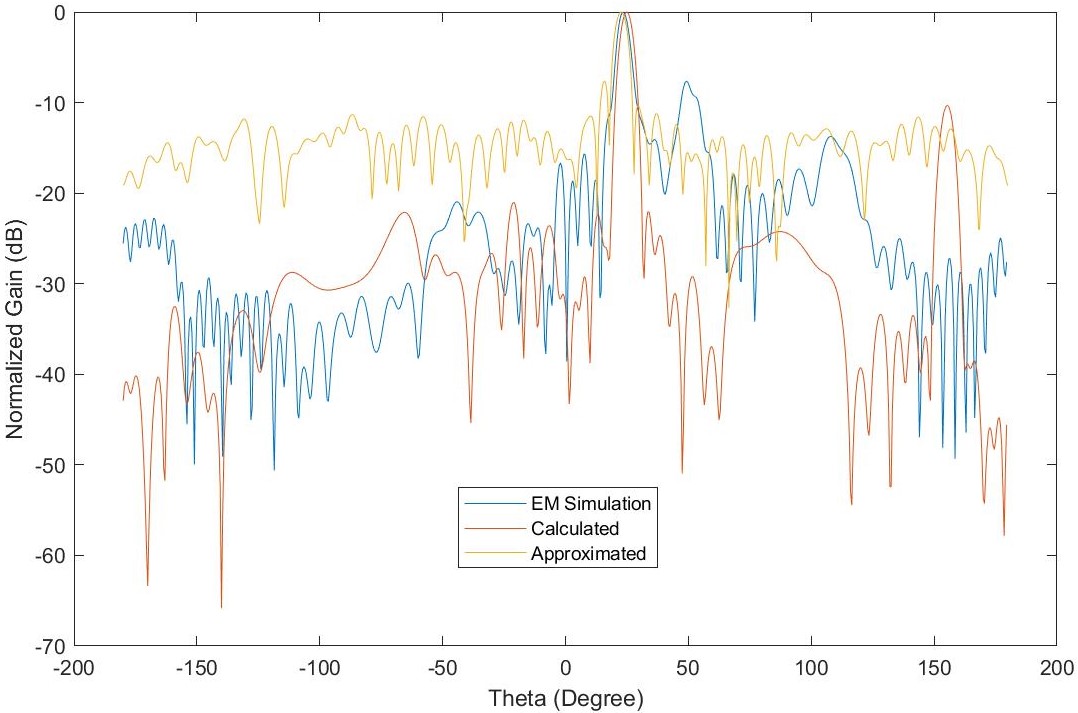}
\caption{Full-wave EM simulated, analytical, and approximated beam patterns when the RIS is configured to reflect the incident wave to the $22.5^\circ$ direction.}
\label{fig:figure_11}
\end{figure}



In the previously discussed Figs.~\ref{fig:figure_10} and~\ref{fig:figure_11}, patterns for the approximated beams designed using \cite{EURASIP_RIS_beams_all}'s Algorithm~$3$ are included, and compared with equivalent analytically computed ones as well as beam patterns obtained by full-wave EM simulations of the fabricated unit cells. As it can be seen, the gain of each beam pattern is normalized in the dB scale, i.e., the main lobe's gain is $0$~dB. It can be observed that the approximated beam patterns can achieve effective focusing at the desired pointing angles since the beam widths at these angles almost coincide among all three cases. The SLL levels of the approximated beams are between $6.5$ and $7$ dB, and the overall beam profiles are flatter and with SLLs placed more closely to the main lobe, as compared to full-wave EM simulations and theoretical calculations for the fabricated responses. All in all, the results in these two figures indicate that, while arbitrary discrete configurations can be produced to perform beam steering at desired angles, more accurate approximation approaches and algorithms are required to reduce reflections at the SLLs. In future work, we plan to extend the approach in \cite{EURASIP_RIS_beams_all} to consider tunable tradeoffs between the gains of the main lobe and those of the SLLs.


\section{Conclusion}
A wideband RIS design operating at the subTHz frequency band was presented in this paper. The proposed design comprised PTCD-based unit cells, each being integrated with an RF switch for realizing a $1$-bit phase shift. The conducted full-wave EM simulations showcased that the designed RIS has excellent beam scanning ability. Directional beams in different direction angles were realized by modifying the $1$-bit phase distribution of the RIS structure.

To deal with the increased complexity of extremely large RIS/array full-wave EM simulations, we presented a fast approximate beam shaping approach that, based on a projection gradient descent technique, approximates ideal continuous reflection patterns to obtain beam patterns of arbitrary large realistic RISs with unit cells of effectively discrete phase shifts. It was demonstrated that the proposed beam shaping approximation technique exhibits sufficient agreement with equivalent full-wave EM simulations in the main lobes while requiring further refinement to match the SLL behavior. The latter constitutes our future research direction which will pave the way for efficient designs of static as well as dynamic beam codebooks for extremely large RIS in subTHz and THz bands.

\section*{Acknowledgments}
This work was supported by the UK Royal Society with grant number ${\rm IES}\! \ \!{\rm R}2\! \ \!212064$ and the Smart Networks and Services Joint Undertaking (SNS JU) under the European Union's Horizon Europe research and innovation programme under Grant Agreement No 101097101, including top-up funding by UK Research and Innovation (UKRI) under the UK government’s Horizon Europe funding guarantee.

\bibliographystyle{IEEEtran}
\bibliography{IEEEabrv,references.bib}
\end{document}